\documentclass[reprint,twocolumn,aps,prl,a4paper,superscriptaddress,showpacs,preprintnumbers,amsmath,amssymb]{revtex4}
\usepackage{CJK} 
\usepackage{color}
\usepackage[colorlinks,linkcolor=blue,anchorcolor=blue,citecolor=blue]{hyperref}
\usepackage{graphicx}
\usepackage{dcolumn}
\usepackage{bm}
\usepackage{ulem}
\usepackage{url}
\usepackage{hyperref}

\begin{document}
\bibliographystyle{unsrt}

\title{Towards the full realization of the RIBLL2 beam line at the HIRFL-CSR complex}
 
\author{Bao-Hua Sun}\thanks{Corresponding author: bhsun@buaa.edu.cn}
\affiliation{School of Physics and Nuclear Energy Engineering, Beihang University, Beijing 100191, China}
\affiliation{International Research Center for Nuclei and Particles in the Cosmos, Beijing 100191, China}
\author{Jian-Wei Zhao} 
\affiliation{School of Physics and Nuclear Energy Engineering, Beihang University, Beijing 100191, China}
\author{Xue-Heng Zhang}  
\affiliation{Institute of Modern Physics, Chinese Academy of Sciences, Lanzhou 730000, China}
\author{Li-Na Sheng}
\affiliation{Institute of Modern Physics, Chinese Academy of Sciences, Lanzhou 730000, China}
\author{Zhi-Yu Sun}
\affiliation{Institute of Modern Physics, Chinese Academy of Sciences, Lanzhou 730000, China}
\author{Isao Tanihata}
\affiliation{School of Physics and Nuclear Energy Engineering, Beihang University, Beijing 100191, China}
\affiliation{International Research Center for Nuclei and Particles in the Cosmos, Beijing 100191, China}
\author{Satoru Terashima}
\affiliation{School of Physics and Nuclear Energy Engineering, Beihang University, Beijing 100191, China}
\affiliation{International Research Center for Nuclei and Particles in the Cosmos, Beijing 100191, China}
\author{Yong Zheng}
\affiliation{Institute of Modern Physics, Chinese Academy of Sciences, Lanzhou 730000, China}
\author{Li-Hua Zhu}
\affiliation{School of Physics and Nuclear Energy Engineering, Beihang University, Beijing 100191, China}
\author{Li-Min Duan}
\affiliation{School of Physics and Nuclear Energy Engineering, Beihang University, Beijing 100191, China}
\author{Liu-Chun He}
\affiliation{School of Physics and Nuclear Energy Engineering, Beihang University, Beijing 100191, China}
\author{Rong-Jiang Hu}
\affiliation{School of Physics and Nuclear Energy Engineering, Beihang University, Beijing 100191, China}
\author{Guang-Shuai Li}
\affiliation{School of Physics and Nuclear Energy Engineering, Beihang University, Beijing 100191, China}
\author{Wen-Jian Lin}
\affiliation{School of Physics and Nuclear Energy Engineering, Beihang University, Beijing 100191, China}
\author{Wei-Ping Lin}
\affiliation{Institute of Modern Physics, Chinese Academy of Sciences, Lanzhou 730000, China}
\author{Chuan-Ye Liu}
\affiliation{School of Physics and Nuclear Energy Engineering, Beihang University, Beijing 100191, China}
\author{Zhong Liu}
\affiliation{Institute of Modern Physics, Chinese Academy of Sciences, Lanzhou 730000, China}
\author{Chen-Gui Lu}
\affiliation{School of Physics and Nuclear Energy Engineering, Beihang University, Beijing 100191, China}
\author{Xin-Wen Ma}
\affiliation{School of Physics and Nuclear Energy Engineering, Beihang University, Beijing 100191, China}
\author{Li-Jun Mao}  
\affiliation{School of Physics and Nuclear Energy Engineering, Beihang University, Beijing 100191, China}
\author{Yi Tian}
\affiliation{School of Physics and Nuclear Energy Engineering, Beihang University, Beijing 100191, China}
\author{Feng Wang}
\affiliation{School of Physics and Nuclear Energy Engineering, Beihang University, Beijing 100191, China}
\author{Meng Wang}
\affiliation{School of Physics and Nuclear Energy Engineering, Beihang University, Beijing 100191, China}
\author{Shi-Tao Wang}
\affiliation{School of Physics and Nuclear Energy Engineering, Beihang University, Beijing 100191, China}
\author{Jia-Wen Xia}
\affiliation{Institute of Modern Physics, Chinese Academy of Sciences, Lanzhou 730000, China}
\author{Xiao-Dong Xu}
\affiliation{School of Physics and Nuclear Energy Engineering, Beihang University, Beijing 100191, China}
\author{Hu-Shan Xu}
\affiliation{Institute of Modern Physics, Chinese Academy of Sciences, Lanzhou 730000, China}
\author{Zhi-Guo Xu}
\affiliation{Institute of Modern Physics, Chinese Academy of Sciences, Lanzhou 730000, China}
\author{Jian-Cheng Yang}
\affiliation{Institute of Modern Physics, Chinese Academy of Sciences, Lanzhou 730000, China}
\author{Da-Yu Yin}
\affiliation{Institute of Modern Physics, Chinese Academy of Sciences, Lanzhou 730000, China}
\author{You-Jin Yuan}
\affiliation{Institute of Modern Physics, Chinese Academy of Sciences, Lanzhou 730000, China}
\author{Wen-Long Zhan}
\affiliation{Institute of Modern Physics, Chinese Academy of Sciences, Lanzhou 730000, China}
\author{Yu-Hu Zhang}
\affiliation{Institute of Modern Physics, Chinese Academy of Sciences, Lanzhou 730000, China}
\author{Xiao-Hong Zhou}
\affiliation{Institute of Modern Physics, Chinese Academy of Sciences, Lanzhou 730000, China}

\date{\today}

\begin{abstract}  
\end{abstract} 


\maketitle

More than 99\% of the mass in the visible universe -- 
the material that makes up ourselves, our planet, stars -- is in the atomic nucleus. 
Although the matter has existed for billions of years, only over the past few decades have we had 
the tools and the knowledge 
necessary to get a basic understanding of the structure and dynamic of nuclei. 
Nuclear physicists around the world have made tremendous strides 
by initiating a broad range of key questions that can be best attacked with various experimental probes at different beam energies.
Moreover, through these efforts, we have gained access to the origin of elements and the 
nucleosynthesis processes that were and still are shaping the world we are living in.

The energy region at around 300 MeV gives rise to the so-called energy window for nuclear structure studies. 
At this energy range, the distortion effects on the projectile wave functions are relatively small due to the weak strength of the scalar-isoscalar interaction, 
which further suppresses the multistep processes in the nuclear reaction mechanism. This brings advantages in studying nuclear spin and isospin excitations~\cite{Osterfeld1992Rev.Mod.Phys.491} and nucleon density distribution of very exotic nuclei characterized by short lifetimes and very different isospins from the stable ones~\cite{Tanihata2013PPNP215,Meng2006Prog.Part.Nucl.Phys.470}. 
Quantitive investigations in the two topics can yield precision information 
 on the weak interaction processes and on how protons and neutrons are distributed in atomic nuclei.
They play important roles not only in nuclear physics but also in astrophysics for stellar events such as supernovae explosions. 

Experimentally, such investigations are closely linked to the availability of separators and spectrometers to select and identify the rare isotopes of interest at relativistic energies of around 300 MeV/nucleon (about 65\% of the speed of light). Among all the separators operating at energies more than300 MeV/nucleon worldwide, 
the Second Radioactive Ion Beam Line in Lanzhou (RIBLL2), one of the key components in the
Heavy Ion Research Facility in Lanzhou (HIRFL-CSR)~\cite{HIRFL2002NIMA} at IMP, China, is unique to have an asymmetric double achromatic configuration. 
 
\begin{figure}[htbp]\noindent
\centering
\includegraphics[width=0.45\textwidth]{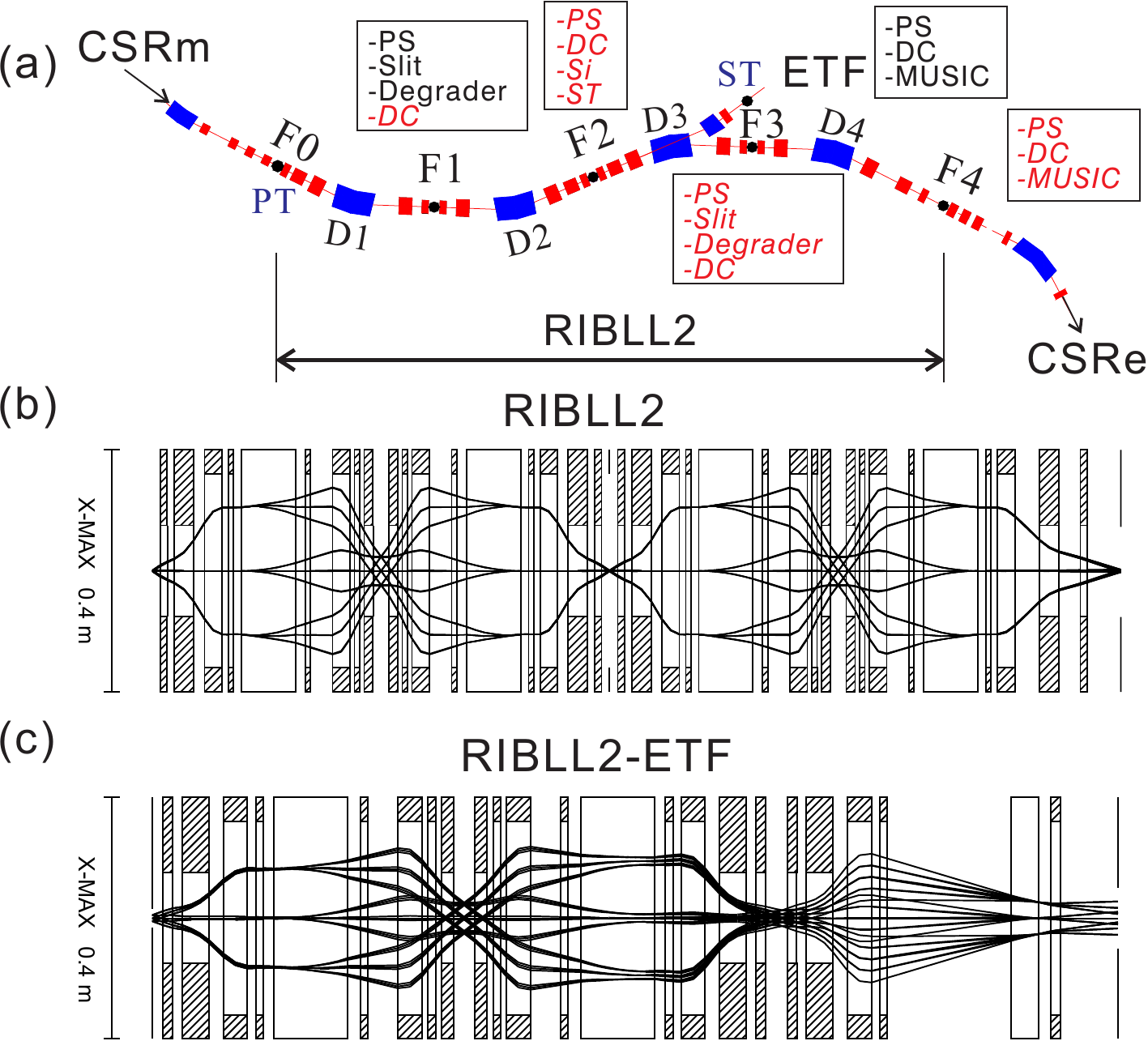} 
 \caption{(Color online) (a) A schematic layout of the RIBLL2 beam line. The labels Dn and Fn indicate the positions of dipole magnets and foci, respectively. 
PT and ST represent the primary target, reaction target, respectively.  
At each focus, PS (plastic scintillator), DC (drift chamber), Si (Silicon detector), MUSIC (multi-sampling ionization chamber), slit and degrader
indicate the types of detectors or devices that have been installed (black letters) or will be installed (red letters).  
Horizontal beam trajectories are calculated with the first-order optics by GICOSY for the full RIBLL2 (F0 to F4) (b) and RIBLL2-ETF (c). 
$^{18}$O$^{8+}$ at 400 MeV/nucleon with a momentum dispersion of $\pm$2\% was used in both calculations. 
The beam emittance of 20$\pi$ and 75$\pi$ mm mrad
was used in the calculation of  RIBLL2 (b) and RIBLL2-ETF (c) configurations, respectively. 
 } \label{fig.1}
\end{figure}

RIBLL2  was constructed in 2007 connecting the synchrotron cooler storage main ring (CSRm)  and  the experimental storage ring (CSRe) in the HIRFL-CSR complex.  
It has been utilized to deliver radioactive isotopes into the CSRe for mass measurements~\cite{CSR2016}. Yet its 
full potential as an individual experimental terminal has not been explored.
The schematic layout of RIBLL2 and external target facility (ETF)  is shown in Fig.~\ref{fig.1}(a).

RIBLL2 has four independent sections, each consisting of a 25$^\circ$ dipole magnet and a set of quadrupole magnets before and after the dipole to fulfill first-order focusing conditions.  Additional 8 hexapole and 4 octupole magnets are equipped for higher-order corrections. 
The whole separator is about 55 meters long, while the first half (F0-F2) and the second half (F2-F4) are about 26 and 29 meters, respectively.  
Shown in Fig.~\ref{fig.1}(b) is the first order ion-optics in the horizontal plane calculated by GICOSY~\cite{GICOSY}.  
The foci at F2 and F4 are fully achromatic.  
The first half (F0-F2) is a mirror-symmetric system, and  can realize a point-to-point image. 
The profile of the second half (F2-F4) is almost reversed to the first half to further purify rare-isotope beam (RIB). 
A maximum dispersion  of 1.169 m/\% can be reached at the dispersive planes F1 and F3. 
Considering the magnification factor of -0.48 from F0 to F1 (or from F2 to F3), the momentum resolution power of 1200 can be expected for a beam size at half width of 1 mm.  

Here we report the recent decisive progress in particle identification for relativistic heavy-ion fragments using RIBLL2. 
About 80 isotopes have been produced and identified unambiguously. 
For the first time, 
we demonstrated that in combination with the external target terminal the first half of RIBLL2 
is able to deliver, separate and identify rare isotopes with the proton number $Z \textless 30$. 

In the experiments, heavy ions were extracted from CSRm to bombard a primary target (PT) placed at the entrance of RIBLL2. 
Rare nuclei of interest were then separated in flight with the first half of RIBLL2 (F0-F2) and delivered to ETF. 
The magnetic rigidity ($B\rho$) of RIB was finely selected by a pair of horizontal passive slits at F1.  
Figure~\ref{fig.1} (c) shows the profile in the horizontal plane of the separator up to ETF.  
A spot size of typically about 5 mm (FWHM), and an emittance of less than about 100 $\pi$mm mrad have been obtained at 
ETF. 

Particle identification with respect to nuclear mass number $A$ and mass-to-charge ratio $A/Q$ is achieved by coincidence measurements of the energy
loss ($\Delta E$) at ETF, $B\rho$ analysis at the dispersive focal plane F1, and the time-of-flight (TOF) determination by two plastic scintillators placed at F1 and ETF. 
$\Delta E$ is measured typically with a large-area stack silicon detector  or a multiple-sampling ionization chamber
(MUSIC)~\cite{ETF-MUSIC1}.

\begin{figure}[htbp!]\noindent
\centering 
\includegraphics[width=0.4\textwidth]{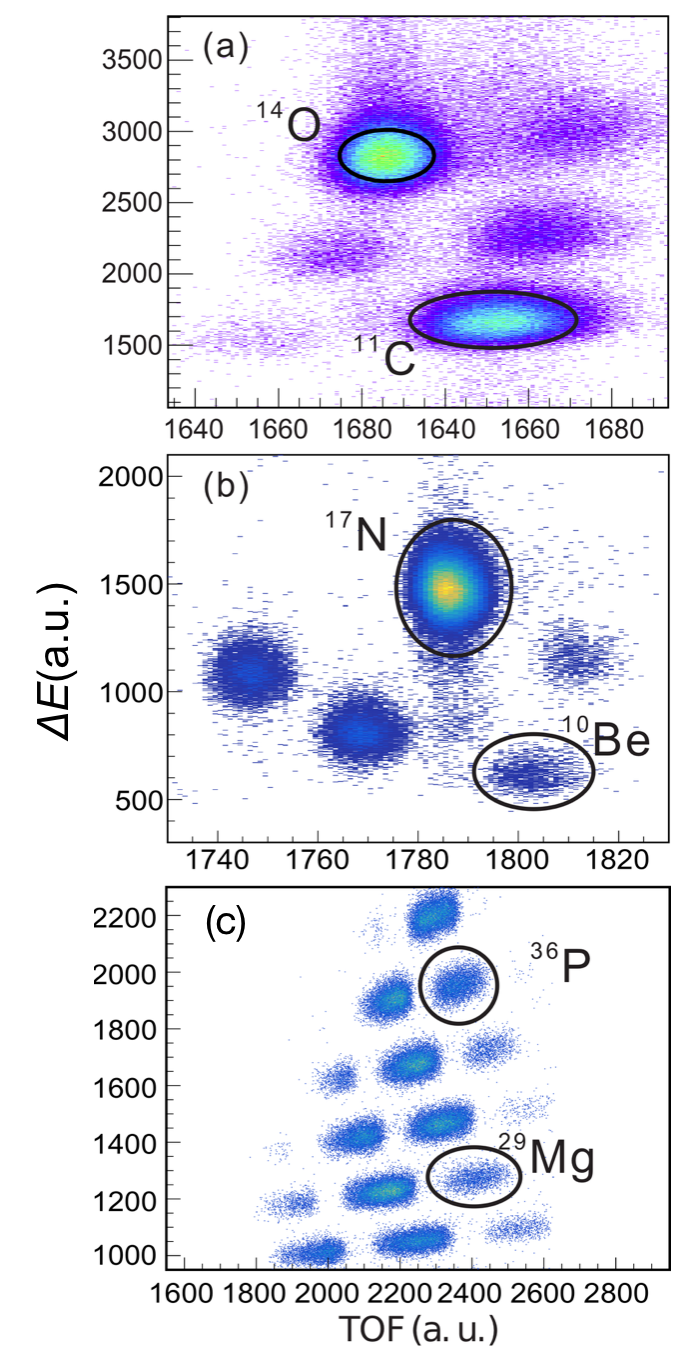}
 \caption{(Color online) Typical particle identification plots of the $^{16}$O, $^{18}$O and $^{40}$Ar fragments
 in three experiments performed in 2013 (a), 2015 (b) and 2016 (c), respectively.  The same ion species groups together in the TOF-$\Delta E$ scatter plot. 
Labelled are several selected identified isotopes. 
The energy loss $\Delta E$ in Fig. (a)-(b) and Fig. (c) were measured with a large-area stack silicon detector and 
the MUSIC, respectively. 
 }\label{fig.pid}
\end{figure}

Figure 2 summarizes the continuously marked progress towards a better isotopic identification in the past years.  
They correspond to the cocktail beams produced by 360 MeV/nucleon $^{16}$O on a 15mm-thick Be target, 400 MeV/nucleon $^{18}$O on a 30 mm-thick Be target, and 320 MeV/nucleon $^{40}$Ar on a 10 mm-thick Be target, respectively. 
This achievement of the isotopic resolution is mainly attributed to the improvement in the beam ion optics and the detector system of
both timing and energy loss determination. In total 77 isotopes were produced and identified via the fragmentations of  $^{18}$O and $^{40}$Ar  (see Supplementary material Table I). In particular, Fig.~2(c) corresponds to the $B\rho$ setting centered in $^{26}$Na (see Supplementary material Table II for the relative production yields). 
The experimental results show that the first half of RIBLL2 is already capable of identifying all ions with $Z \textless 30$. 
 
Due to the inevitable channeling effect in silicon detectors (see in particular the $^{17}$N case in  Fig.~2(b)),  
the MUSIC detector is preferred for relativistic heavy ions. 
A charge resolution of 0.12 ($\sigma$) has been achieved  
as demonstrated in Fig.~2(c), and it does not change for the range of elements measured here. 

The intrinsic TOF resolution ($\sigma$) from detectors is determined as 80 ps~\cite{Lin2017CPC}. 
The TOF-start detector viewed by one PMT at F1 is a plastic scintillator with a cross-section of 100 mm $\times$ 100 mm and a thickness of 3 mm.  
Such relatively large size is required to cover the large beam emittance at the F1 dispersive plane.   
The TOF-stop detector at ETF has a dimension of 50 mm $\times$ 50mm $\times$ 3 mm, and is viewed by two fast PMTs
positioned at both ends of the plastic scintillator. Its intrinsic time resolution is determined to be about 30 ps ($\sigma$)~\cite{Zhao2016NIMA,Lin2017CPC}. 

The initial $B\rho$ or momentum spread from the fragmentation process is the dominant source for 
the final mass resolution.  A detector with a position resolution of less than 0.5 mm at F1 
can make a precise $B\rho$ measurement possible and this can further improve the mass resolution. 
 
A full commissioning RIBLL2 requires a complete detector system for both the beam monitoring and 
the ion tracing measurements (see Fig.~1(a)), which brings advantages in a 
better isotopic purification due to more freedom to manipulate magnetically the secondary ions and in better isotopic identification ability 
due to a much longer TOF path length. The total TOF length is 42 meters from F1 to F4. This is by 16 meters longer than that of RIBLL2-ETF, 
thus a much better separation quality can be expected for heavier nuclear system. 

In particular, the unique asymmetric arrangement provides a natural way to operate RIBLL2 in the separator-spectrometer mode. 
The first stage (half) of RIBLL2 can be used to produce and separate RIB with the $B\rho-\Delta E-B\rho$ method (see Fig.~1(a)-(b)). 
An energy degrader for $\Delta E$ analysis will be installed at F1.   
A secondary reaction target can be placed at the achromatic focus F2.
Then the second stage (F2-F4) can work as a high-resolution zero-degree spectrometer, to analyze and identify further 
the outgoing reaction products with the standard TOF-$B\rho$-$\Delta E$ on an event-by-event basis.  
Another energy degrader can be placed at F3 to further purify the RIBs.  
Such a full realization of the RIBLL2 beam line is going to open new opportunities for addressing many interesting and important nuclear structure problems at around 300 MeV/nucleon, 
one of which is a new direct method to weigh masses of most neutron-rich atomic nuclei~\cite{Sun2015FronPhys10}.   
It should be also mentioned that the development of RIBLL2 would provide a very valuable experience 
to design the high-energy fragment separator at the High Intensity heavy ion Accelerator Facility (HIAF)~\cite{HIAF2013}
in China. 
 
\textbf{Conflict of interest}

    The authors declare that they have no conflict interests.  
  
\textbf{Acknowledgments}

The authors thank the staffs in the accelerator division of
IMP for providing stable beams, and would like to express their congratulation on the 60 years' anniversary of IMP, CAS.  
This work has been supported by the National Key R\&D program of China (Grant No. 2016YFA0400504) and 
the National Natural Science Foundation of China (Grant No. 11475014, and No. 11235002).



\end{document}